\documentclass[aps,prl,twocolumn,superscriptaddress,showpacs,floatfix]{revtex4}

\usepackage{graphicx}
\usepackage{color}

\usepackage{bm}
\usepackage{amssymb}
\usepackage{amsmath}
\newcommand{\be}{\begin{equation}}
\newcommand{\ee}{\end{equation}}

\begin{document}
\title{Statistics of small scale vortex filaments in turbulence}
\author{Luca Biferale} \affiliation{Dept. Physics and INFN, University
  of Tor Vergata, Via della Ricerca Scientifica 1, 00133 Rome,
  Italy.\\International Collaboration for Turbulence Research.}
\author{Andrea Scagliarini} \affiliation{Dept. Physics and INFN,
  University of Tor Vergata, Via della Ricerca Scientifica 1, 00133
  Rome, Italy.\\International Collaboration for Turbulence Research.}
\author{Federico Toschi} \affiliation{Department of Physics and
  Department of Mathematics\\and Computer Science and J. M. Burgers
  Centre for Fluid Dynamics,\\Eindhoven University of Technology, 5600
  MB Eindhoven, The Netherlands.\\International Collaboration for
  Turbulence Research.}

\begin{abstract}
  We study the statistical properties of coherent, small-scales,
  filamentary-like structures in Turbulence.  In order to follow in
  time such complex spatial structures, we integrate Lagrangian and
  Eulerian measurements by seeding the flow with light particles.  We
  show that light particles preferentially concentrate in small
  filamentary regions of high persistent vorticity (vortex
  filaments). We measure the fractal dimension of the attracting set
  and the probability that two particles do not separate for long time
  lapses.  We fortify the signal-to-noise ratio by exploiting
  multi-particles correlations on the dynamics of bunches of
  particles. In doing that, we are able to give a first quantitative
  estimation of the vortex-filaments life-times, showing the presence
  of events as long as the integral correlation time.  The same
  technique introduced here could be used in experiments as long as one
  is capable to track clouds of bubbles in turbulence for a relatively
  long period of time, at high Reynolds numbers; shading light on the
  dynamics of small-scale vorticity in realistic turbulent flows. 

\end{abstract}
\pacs{47.27.-i, 47.10.-g, 47.11.-j} 

\date{\today}
\maketitle
Vorticity dynamics, in general, and vortex filaments, in particular,
have been the subject of many theoretical, phenomenological and
experimental studies from the turbulent community, both applied and
theoretical \cite{saffman,frisch,moffat,greg}.  The transport of
particulate by fluid flows is an ubiquitous phenomenon in nature and
in industrial applications alike.  According to their inertia
properties particles respond differently to fluctuations of the
advecting -Eulerian- velocity field producing locally non homogeneous
concentration, a phenomenon dubbed as preferential concentration
\cite{tb2009,eaton}. As an example of such differential response, it
has been clearly shown in the past that heavy particles tend to be
expelled from high vorticity regions, while light particles tend to
concentrate in regions where vorticity is higher \cite{tb2009}.
Thanks to their strong property to concentrate in high vorticity
regions, very light particles (i.e. small bubbles in water) have been
used to visualize small scale vortex filaments
\cite{douady:1991,bonn:1993}.  The strong tendency of light particles
to concentrate in vortex filaments has been quantified also measuring
the fractal dimensions of their dynamical attractor \cite{jimenez}.
Similar phenomena, based on complex response of microscopic hydrogen
particles in quantum fluids, have also been exploited recently to
visualize quantized vortices \cite{nature}.

In the present work we will focus on the dynamics of light particles
and will study in detail the connection between their dynamics and the
one of small scale vortex filaments \cite{moffat}.

The data used for this study come from Direct Numerical Simulation
(DNS) of 3D fully periodic Navier-Stokes eqs plus particles.  Indeed,
together with the Eulerian field we integrated the Lagrangian
evolution of particles by mean of one of the simplest, yet nontrivial,
model of dilute, passively advected, suspensions of spherical
particles as described in Refs.~\cite{maxey:1983,auton:1988} :
\begin{equation} {{{\rm d}\bm x} \over {{\rm d} t}}=\bm v\,,\qquad
  {{{\rm d}\bm v} \over {{\rm d} t}} =\beta \, {{{\rm D}\bm u} \over
    {{\rm D} t}} + {1\over {\tau_p}}(\bm u -\bm v)\,,
\label{eq:dynamics}
\end{equation}
In the above equation $\bm x(t)$ and $\bm v(t)$ denote the particle
position and velocity respectively, $\tau=a^2/(3\beta \nu)$ is the
particle response time, $a$ is the particle radius and
$St=\tau_p/\tau_{\eta}$ is the Stokes number of the particle and
$\beta= 3 \rho_f / (\rho_f + 2 \rho_p)$ is related to the contrast
between the density of the particle, $\rho_p$, and that of the fluid,
$\rho_f$. The incompressible fluid velocity $\bm u(\bm x,t)$ evolves
according to the Navier-Stokes equations\ :
\begin{equation} {\rm D}_t \bm u \equiv \partial_t \bm u + \bm u \cdot \bm
  \nabla \bm u = -\bm \nabla p +\nu \Delta \bm u + \bm f
  \,.\label{eq:ns}
\end{equation} 
where $p$ denotes the pressure and $\bm f$ an external forcing
injecting energy at a rate $\epsilon=\langle \bm u\cdot \bm
f\rangle$. Eqn.~(\ref{eq:ns}) is evolved by means of a pseudo-spectral
code for the fluid part with a second order Adams-Bashforth
integrator, also used for the --millions of-- particles evolving
according to the dynamics (\ref{eq:dynamics}), where the fluid
velocity at the particle position, $\bm u(\bm x(t),t)$, was obtained
by means of a tri-linear interpolation \cite{mazzitelli:2003,
  biferale:2004g}.  Energy was injected maintaining constant the
spectral content of the first two shells in Fourier space.  We report
data coming from two sets of simulations with $N^3=128^3$ and
$N^3=512^3$ collocation points, corresponding to $\mathrm{Re}_\lambda
\simeq 80$ and $\mathrm{Re}_\lambda \simeq 180$, respectively.  We
focus mainly on very light particles, in the limit of $\rho_p
\rightarrow 0$ ($\beta \rightarrow 3$ ) and on tracers evolving with
the local Eulerian velocity field $\dot{\bm{x}}(t) = {\bm u(\bm
  x(t),t)}$.

Inertial particles are not distributed homogeneously in the volume,
centrifugal forces tends to concentrate light particle inside strong
elliptical regions (with high vorticity) and heavy particles in
hyperbolic regions, typical of intense shear. One thus expects
different temporal correlations between particle trajectories and the
underlying topology of the carrier flow. The local topology of the
Eulerian flow is defined in terms of the symmetric and anti-symmetric
component of the gradients $A_{ij} = \partial_i u_j$, hence there is
an intimate link between the statistics of energy dissipation and/or
enstrophy with particles evolution
\cite{yeung:2007,luthi:2005,guala:2005}.
\begin{figure}[t!]
\includegraphics[width=\hsize]{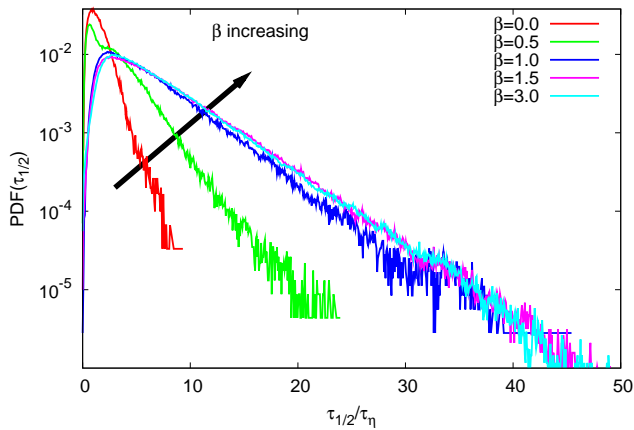}
\caption{PDF of \textit{halving--times} (see text) for various density
  ratios, $\beta$, at fixed $St\sim0.5$, conditioned on the initial
  value of the vorticity magnitude to be larger than a given
  threshold. As $\beta$ increases the PDFs show larger and larger
  tails, suggesting that lighter particles are probing regions of high
  and stable vorticity.}
\label{fig0}
\end{figure}
To give an idea of this effect, we show in Fig. \ref{fig0} the
conditional probability distribution of the halving times,
$\tau_{1/2}$, of the vorticity magnitude along particle
trajectories. The vorticity halving time is defined, given a time $t$,
as the first time lag after which the vorticity has becomes $1/2$
larger or smaller than the initial value, $\tau_{1/2}(t) = min(\tau |
\omega(t+\tau)/\omega(t) = 1 \pm 1/2)$. The PDFs are conditioned in
such a way that the vorticity magnitude at the reference time $t$ is
greater than a given threshold (for the case in Fig. \ref{fig0}, we
chose this value to be $5\omega_{rms}$; indeed the shape of the PDFs
does not change significantly for higher values of the threshold).  As
one can see from the inset of Fig. \ref{fig0}, for a given Stokes, $St
\sim 0.5$, at changing the density contrast, $\beta$, one moves
towards higher and higher probability to observe long halving times,
i.e. light particles tend to live in regions of high and stable
vorticity.

\begin{figure}[t!]
\includegraphics[width=.92\hsize]{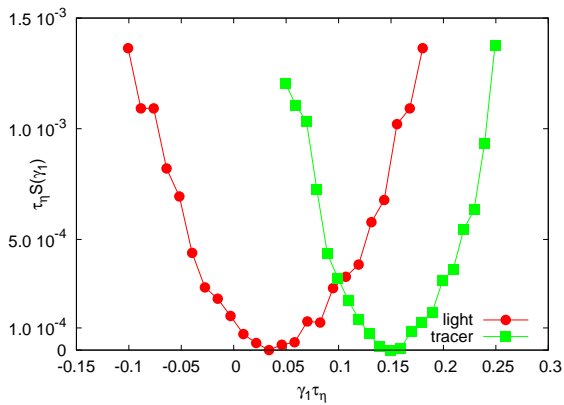}
\caption{Cram\'er entropy for light particles and tracers. The
  function, for the case of light particles, is shifted towards lower
  values (the maximum Lyapunov exponent where $S(\gamma_1)$ attains
  its minimum is smaller than the one for tracers) and shows larger
  tails (thus suggesting a higher degree of intermittency). There is a
  non--negligible fraction of events (left tail of the red curve)
  with couples of particles that do not separate at all (absent in the
  tracers case).}
\label{fig2}
\end{figure}

Stable vortex structures, chaotically advected by turbulent flows,
should therefore play the role of attractive sinks in the
dynamics. Larger their lifetime will be, larger inhomogeneous bubble
distribution will develop.  To quantify better the clustering
properties of light particles, we start from studying the statistics
of the largest Lyapunov exponent, $\lambda_1$.  It is known that
clustering and inertia may affect the whole distribution of Lyapunov
exponents. One would expects that inertia reduces the tendency for
particles pairs to separate. Indeed, a small tendency toward a
reduction of the Lyapunov exponent at increasing the Stokes number has
been reported for heavy particles \cite{bec_et_al_pof:2006}, similar
results have also been obtained in random flows
\cite{duncan-mehlig-prl}. Even more interesting is the study of the
whole probability distribution of the largest finite-time Lyapunov
exponent (FTLE), $\gamma_1(T)$ defined as: $\gamma_1(T) =
\frac{1}{T}\log(R(T,t)/R(0,t))$ where $R(T,t)$ is the separation of
two particles at time $t+T$ starting from a separation $R(0,t)$ at
time $t$.  For large times, the distribution of FTLE is expected to
follow a large-deviation form: $p_T(\gamma_1) \propto
\exp[-TS(\gamma_1)]$, where the Cram\'er function, $S(\gamma_1)$ is a
non-negative convex function vanishing for $\gamma_1=\lambda_1$.  In
Fig. \ref{fig2} we report the Cram\'er function at $T=190\tau_\eta$
for the case of light particles $\beta=3, St\sim 1.2$ compared with
the one obtained for tracers. As one can see there are two remarkable
effects. First, the minimum for the case of light particles is
achieved for a value much smaller than the one for the tracers,
precisely $\gamma_1 \tau_\eta \sim 0.04$ for bubbles and $\gamma_1
\tau_\eta \sim 0.14$ for tracers. Second light particles show a
remarkably high probability to have pairs that do not separate at all,
even for long time lapses, i.e. there are many events in the left tail
of the Cram\'er function which have a negative FTLE. The global
average properties, however, remains chaotic, i.e. with probability
one all couples separate at long times. The Kaplan-Yorke dimension for
the family of light particles shown in Fig. \ref{fig2} is $D_{KY} =
1.3 \pm 0.3$ \cite{calza}.
The observation that such small-scales strong clustering properties are 
correlated to the topology of the flow structures at the same scales
suggests  the possibility to use
light particles to study statistical properties of small-scales vortex
filaments in turbulence.

The identification of small scale vortex filaments is an extremely
daunting task. An analysis of the Eulerian fields would require the
detection of isosurfaces of vorticity (larger than some prescribed
threshold): a method problematic if not for the larger and most
intense structures \cite{jimenez}. Furthermore to analyze the temporal
evolution of vortex filaments one would also need to track three
dimensional structures not only in space, but also to follow their
evolution in time by repeating the same analysis at each Eulerian time
step.
\begin{figure}[t!]
\includegraphics[width=\hsize]{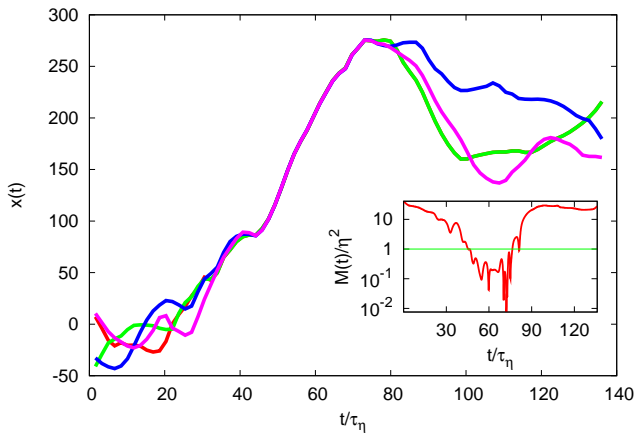}
\caption{Evolution of the trajectory of different light particles
  entering in a small scale vortex filament and then separating
  again. In the inset it is shown the behavior of the momentum of
  inertia of the particles cloud. At entering (exiting) from the
  vortex filament an abrupt --notice the log scale-- decrease (increase) of the signal allows to
  sharply define the life--time of the vortex filament.}
\label{fig1}
\end{figure}
Our original proposal here is to use multi-particles correlation to
extraordinary enhance the signal-to-noise ratio associated to the
identification of vortex filaments. In Fig. \ref{fig1} we show the
trajectories of several light particles which are attracted, in
presence of favorable pressure gradients, into a vortex filament and
that then separate again once the vortex filament disappears. Only the
favorable pressure gradient, which tends to concentrate bubbles in
the vortex core, permits such a strong clustering. Once the vortex
filament breaks down, particles are released and separate explosively.

\begin{figure}[t!]
  \includegraphics[width=\hsize]{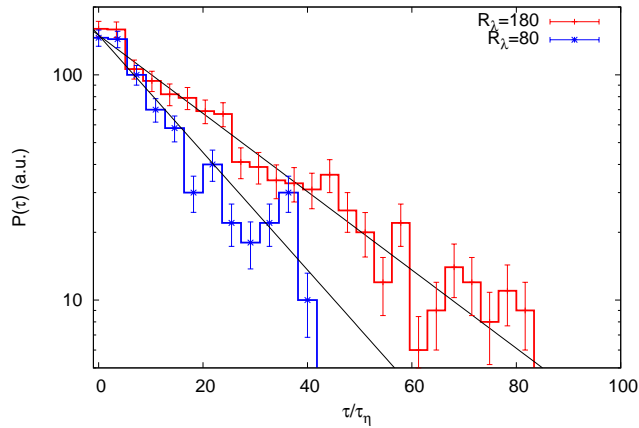}
  \caption{Probability distribution function of the life--time of
    vortex filaments. PDFs at two different Reynolds number,
    $R_{\lambda}=80$ and $R_{\lambda}=180$ respectively: in both cases
    it turns out that the PDF is fairly well fitted by an exponential
    (with a decay factor, for the highest $R_{\lambda}$, of around $20
    \tau_{\eta}$). In the far tail of the PDFs there are events whose
    life--time is of the order of the integral time. Error bars
    reflect the effect of changing the threshold used to determine the
    vortex filament time of life (see text for more details). }
\label{fig3}
\end{figure}

\begin{figure}[t!]
  \includegraphics[width=\hsize]{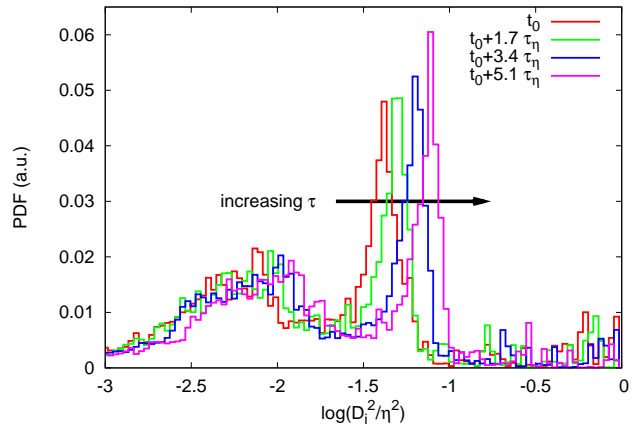}
  \caption{The probability distribution function of the average
    distance (squared) between pairs of particles in a bunch clearly
    showing a fragmentation process: a vortex filament is splitted in
    two and the bunch of particles is splitted in two bunches
    separating in time (the distance between the bunches is associated
    with the moving position of the peak).}
\label{fig4}
\end{figure}

One can define the signal-to-noise (s/n) ratio for an event with $n$
particles in a volume where on average there are $\overline n$ as:
$\mbox{SNR} = {n/{\overline n}}$.  Our goal here is to identify an
observable which is very sensitive to the presence of small scale
vortex filaments.  With such an observable it will be possible to fix
a threshold to define a birth and death time for the vortex filaments.
We now outline our procedure in detail.  We take a snapshot of light
particles configuration (i.e. positions) at a time roughly in the
middle of our numerical simulation. We divide our simulation volume
into small cubes, say of size $4^3$ Kolmogorov scales, $\eta$, and we
look for those volumes with larger particle counts. The particles
residing in a volume will form what we call a bunch. We then consider
the full trajectory of the particles within the several bunches that
we identified. Obviously the number of particles in the different
bunches will not be the same. We identify $M$ bunches with the $i$-th
bunch formed by $N_i$ particles, then we define the center of mass of
the bunch $i$ as $ {\bm x}^i_{cm}(t) = ({1/ {N_i}}) \sum_{j=1}^{N_i}
{\bm x}_j(t) $ and the momentum of inertia for the same bunch of
particles as: $M_i = {1\over {N_i}}\sum_{j=1}^{N_i} \left( {\bm
    x}_j(t) - {\bm x}^i_{cm}(t) \right)^2$.
The physical interpretation is clear, the smaller the momentum of
inertia of the bunch, the closer the particles. The important
observation here is that this quantity is very sensitive to vortex
filaments and displays an extraordinarily high signal/noise ratio (see
inset of Figure \ref{fig1}). To understand this point one has to
consider the fact that we find easily hundreds of particles at
distances smaller than $0.1\eta$, while for an uniform
distribution one would expect to find $1.4\cdot 10^{-3}$ particles per
$\eta^3$. The probability to find a finite number of particles (even
if just a few, say $3 \div 4$) close inside $\eta^3$ is so small that
if this happens it is almost surely associated to the presence of
confining forces keeping the particles close-by (the probability could
be estimate by means of the Poissonian distribution).\\
Another remarkable feature that makes the momentum of inertia of the
bunch an extremely useful quantity to identify vortex filaments is the
rapidity with which light particles in the neighborhood of a forming
vortex filament are attracted into it. In the inset of Fig.
\ref{fig1} one can indeed see that particles initially separated move
closer, remain very close to each other for some time, and then
separate again. The extreme sharpness (rapidity) with which the
momentum of inertia drops and then raises again allows to define the
life-time of the vortex filament in a robust way. The life-time
estimates are not much sensitive to the chosen threshold (in
Fig. \ref{fig1} the threshold has been set to the value 1, changing
the value of the threshold is accounted into the error bars e.g. in
Fig. \ref{fig3}).\\
By analyzing the statistics of all bunches, we can make an histogram
of vortex filaments life--times, allowing for the first time to
assess, in a quantitative way, the statistical properties of these
extreme events. In Fig. \ref{fig3} we show the pdf of vortex filaments
life--times for $Re_\lambda=180$ and $80$ which happens to be an
exponential with a decay rate which can be estimated of the order of
$25 \tau_{\eta}$ and $17\tau_{\eta}$, respectively. As already noticed
in \cite{douady:1991} we do observe events as long as the integral
time $T_L$, that in our simulation was estimated to be of the order of
$50 \tau_{\eta}$.  
Another interesting question is how particles bunches are formed, if
this happens by means of a sudden attraction of close-by particles, or
as the result of a sequence of successive -and rapid- bunch merge. The
same question can be posed regarding the ``decay'' of a vortex
filament. How do particles separate?  To answer this question one can
introduce the following quantity measuring the average distance
between two particles in a bunch:
\begin{equation}
  D^2_i = {2\over {N_i(N_i-1)}} \sum_{j,k=1}^{N_i} \left( {\bm x}_j(t) - {\bm x}_k(t) \right)^2
\end{equation}
Should all particles be close then the histogram of the distances
would be centered around a small unique value with a variance
connected with the bunch size. In the event of a bunch being splitted
into two separating (smaller) bunches, one would expect to see the
appearance of a peak at some finite distance and a shift of the
position of this peak with time. This is exactly the case in
Fig. \ref{fig4} where the pdf of $D^2_i$ for a given bunch is shown
for 4 consecutive instants in time. A clearly visible peak can be
identified and its position is shifting towards larger distances.

One of the most intriguing features of fluid dynamics turbulence is
the presence of long living coherent structures at small scales.  The
quantification of the statistical properties of these Eulerian
structures has always proved to be one of the most difficult
statistical analysis in turbulence due to the extremely low signal to
noise ratio and to the need of following the evolution of structures
for as long as a large scale eddy turnover time.  We showed, for the
first time, that the coherent dynamical properties of clouds of very
light particles (bubbles) can be used to introduce an observable
extremely sensitive to small scale vorticity filaments.  We used this
observable to quantitatively measure the probability distribution of
vortex filaments life--times. We also show that the decay of a vortex
filament into smaller chunks is the way particles get released from
within such structures.  The same observable introduced here could be
used in experiments as long as one is capable to track clouds of
bubbles in turbulence for a relatively long period of time, at high
Reynolds numbers. Vortex dynamics is  of particular importance also to
study possible blow-up of Euler and Navier-Stokes equations \cite{recon} as
well as super-fluid dynamics \cite{koplik}. 
Similar ideas could abridged to other domain to
fortify statistical signals.

{\bf Acknowledgments} We thank the DEISA Consortium (co-funded by the
EU, FP6 project 508830), for support within the DEISA Extreme
Computing Initiative (www.deisa.org).  Data from this study are
publicly available in unprocessed raw format from the iCFDdatabase
(http://cfd.cineca.it).


\end{document}